\documentclass[aps,prx,twocolumn,superscriptaddress,showpacs,floatfix,preprintnumbers]{revtex4-2}

\usepackage[dvipsnames]{xcolor}     
\definecolor{lcolor}{rgb}{0.5,0,0}
\definecolor{citcolor}{rgb}{0,0,1}
\usepackage[breaklinks,colorlinks,urlcolor=blue,citecolor=citcolor,linkcolor=lcolor,linktoc=all]{hyperref}
\usepackage{color}
\usepackage{graphicx}
\usepackage{dsfont}
\usepackage{multirow}
\graphicspath{{./figures/}}
\usepackage[utf8]{inputenc}
\usepackage{amsmath} \usepackage{amssymb}
\usepackage[dvipsnames]{xcolor}      

\allowdisplaybreaks

\makeatletter
\g@addto@macro\bfseries{\boldmath}
\makeatother

\usepackage{tikz}
\usepackage[customcolors]{hf-tikz}
\usepackage{mciteplus}

\usetikzlibrary{arrows,cd,shapes,decorations.pathmorphing,decorations.markings,shadings}
\tikzset{
  big arrow/.style={
    decoration={markings,mark=at position 1 with {\arrow[scale=4,#1]{>}}},
    postaction={decorate},
    shorten >=0.4pt},
  big arrow/.default=blue}

\newcommand{\cint}{\mathrel{C\!\!\!\!\!\!\!\!\:\int}}

\newcommand{\bbone}{\text{\usefont{U}{bbold}{m}{n}1}}
\MakeRobust{\bbone}

\newcommand{\Lbar}{\bar{\Lambda}}

\newcommand{\nf}{N_f}

\newcommand{\pt}{{\vec{p}}}

\newcommand{\st}{{\vec{s}}}

\renewcommand{\vec}{\mathbf}

\renewcommand{\epsilon}{\varepsilon}

\newcommand{\ud}{\mathrm{d}}

\newcommand{\gamE}{\gamma_{\text{E}}}

\usepackage[acronym]{glossaries}
\newacronym{LO}{LO}{leading-order}

\newacronym{QM}{QM}{quark matter}

\newacronym{QCD}{QCD}{quantum chromodynamics}

\newacronym{HTL}{HTL}{Hard-Thermal-Loop}

\newacronym{UV}{UV}{ultraviolet}

\newacronym{IR}{IR}{infrared}

\begin{document}

\title{Quark matter at four loops: hardships and how to overcome them
}

\preprint{HIP-2025-01/TH}

\author{Aapeli K\"arkk\"ainen}
\affiliation{Department of Physics and Helsinki Institute of Physics,
P.O.~Box 64, FI-00014 University of Helsinki, Finland}
\author{Pablo Navarrete}
\affiliation{Department of Physics and Helsinki Institute of Physics,
P.O.~Box 64, FI-00014 University of Helsinki, Finland}
\author{Mika Nurmela}
\affiliation{Department of Physics and Helsinki Institute of Physics,
P.O.~Box 64, FI-00014 University of Helsinki, Finland}
\author{Risto Paatelainen}
\affiliation{Department of Physics and Helsinki Institute of Physics,
P.O.~Box 64, FI-00014 University of Helsinki, Finland}
\author{Kaapo Seppänen}
\affiliation{Department of Physics and Helsinki Institute of Physics,
P.O.~Box 64, FI-00014 University of Helsinki, Finland}
\author{Aleksi Vuorinen}
\affiliation{Department of Physics and Helsinki Institute of Physics,
P.O.~Box 64, FI-00014 University of Helsinki, Finland}

\begin{abstract}

\noindent Knowledge of the pressure of cold and dense quark matter (QM) is known to significantly constrain the equation of state of neutron-star matter, a quantity playing a key role in deciphering the stars' internal structure. In this work, we make important progress towards determining the last unknown contribution to the pressure at Next-to-Next-to-Next-to-Leading Order (N3LO) in the strong coupling constant $\alpha_s$, available through the sum of all four-loop vacuum diagrams of dense Quantum Chromodynamics. We demonstrate the cancellation of both the covariant gauge parameter and all infrared (IR) divergences in the sum, showcasing the effective-field-theory paradigm in action. For the remaining IR-finite four-loop integrals, we demonstrate the efficacy of the dense Loop Tree Duality method --- a novel numerical framework for multiloop calculations in thermal field theory. Together, our results show that completing the N3LO pressure of cold QM is no longer merely possible but for the first time within reach.

\end{abstract}

\maketitle

\section{Introduction}

The equation of state (EoS) of cold and dense deconfined Quantum Chromodynamics (QCD) matter, quark matter (QM) in short, is a crucial ingredient in the model-agnostic inference of the neutron-star-matter EoS, for which it provides a crucial high-density constraint \cite{Annala:2021gom,Komoltsev:2021jzg,Annala:2023cwx}. It is thus hardly surprising that considerable effort has been placed into improving a classic calculation by Freedman and McLerran from nearly 50 years ago, valid up to order $\alpha_s^2$ in the strong coupling constant $\alpha_s=g^2/(4\pi)$ \cite{Freedman:1976ub}. This landmark result has since been complemented by strange-quark-mass effects \cite{Kurkela:2009gj,Gorda:2021gha,Fernandez:2024ilg} as well as the purely soft \cite{Gorda:2018gpy,Gorda:2021kme,Gorda:2021znl,Fernandez:2021jfr} and hard-soft \cite{Gorda:2023mkk} contributions at $O(\alpha_s^3)$, with the hard scale given by the baryon chemical potential $\mu_\text{B}$ and the soft one by the in-medium screening mass $m_\text{E}\sim g\mu_\text{B}$. What is, however, still missing from a full Next-to-Next-to-Next-to-Leading Order (N3LO) result is the numerically dominant contribution of the hard scale, given by the sum of all four-loop vacuum diagrams of QCD. The completion of this calculation is expected to dramatically improve the accuracy of the result \cite{Gorda:2023mkk}.

The technical challenge associated with evaluating four-loop diagrams in a thermal setting is formidable, and only a handful of examples of successfully completed graphs exist. This is largely due to many crucial tools of vacuum ($T=\mu_\text{B}=0$) perturbation theory, such as four-dimensional Integration-by-Parts (IBP), being weakened or absent due to the effective breaking of Lorentz invariance \cite{Osterman:2023tnt}. At nonzero temperature $T$ but vanishing chemical potentials, fully evaluated four-loop diagrams include those appearing in $\phi^4$ theory \cite{Gynther:2007bw} as well as one single diagram in QCD \cite{Gynther:2009qf}, proportional to the maximal power of the number of fermion flavors, $N_f^3$, while in the opposite limit of high $\mu_\text{B}$ and $T=0$ even fewer successfully determined four-loop integrals exist \cite{Gorda:2022zyc,Navarrete:2024zgz}.

In the cold and dense limit, the established machinery of perturbative thermal field theory consists of the so-called cutting rules \cite{Ghisoiu:2016swa}, which unfortunately suffer from the inadvertent splitting of infrared (IR) convergent entities to multiple IR-sensitive integrals of varying dimensions, complicating calculations especially in the limit of massless quarks. This has motivated a search for alternative techniques, of which a particularly promising candidate was recently introduced in \cite{Navarrete:2024zgz}. This method carries the name dense Loop Tree Duality (dLTD) and amounts to a highly automatized numerical framework, inspired by recent advances in the high-performance evaluation of scattering amplitudes in collider physics (see, e.g., \cite{Catani:2008xa,Capatti:2019ypt,Capatti:2020xjc}). In dLTD, one begins by analytically performing all temporal momentum integrals of the diagram, which results in a locally finite representation of the integrand, to be numerically integrated with Monte-Carlo tools. Achieving local finiteness in $D=4$ dimensions requires a subtraction procedure for the original IR and UV poles, of which the latter can be handled using the well-established methods of vacuum field theory \cite{Bogoliubov:1957gp,Hepp:1966eg,Zimmermann:1969jj,Capatti:2019edf,Capatti:2022tit}.

In this paper, we take first steps towards evaluating not only a few isolated integrals but the set of all four-loop vacuum graphs of cold and dense massless QCD. First, we utilize a finite-density generalization of the canonicalization procedure of \cite{Navarrete:2024ruu} to reduce the diagrams to integrals we refer to as masters, finding six group-theory-invariant sectors and witnessing a full cancellation of the covariant gauge parameter $\xi$. Next, we inspect the IR-sensitive diagrams containing multiple quark loops, identify the subspace of IR-divergent integrals mapping to the Hard Thermal Loop (HTL) soft effective theory, and witness another full cancellation of the IR poles against HTL structures. Finally, we inspect a representative set of the remaining IR-finite integrals, demonstrating that the numerical results obtained with the dLTD method are in full agreement with an independent semi-analytic calculation, where we employ the traditional methods.

\begin{figure*}[t!]
    \includegraphics[width=\textwidth]{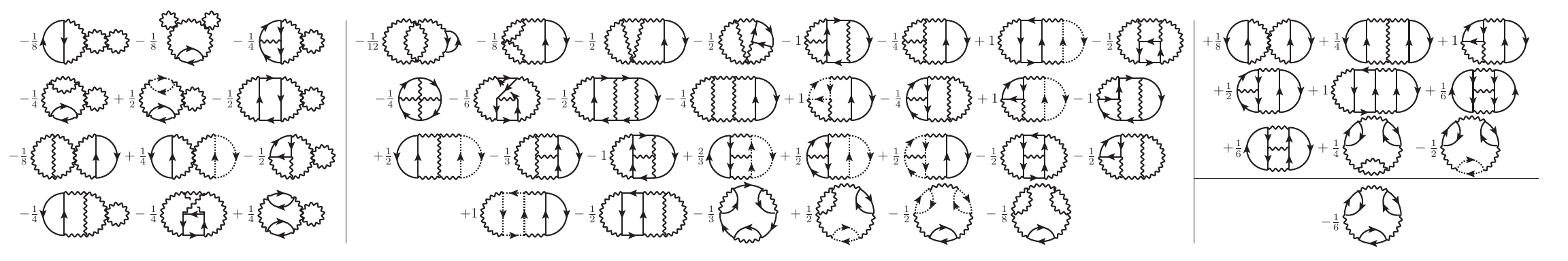}
\caption{All four-loop vacuum diagrams of QCD with at least one quark loop, shown with their signs and symmetry factors.}
\label{fig:diags}
\end{figure*}

\section{Organization of the diagrams}

We consider the thermodynamic pressure $p$ of deconfined QCD matter, keeping the numbers of colors $N_c$ and (massless) quark flavors $N_f$ arbitrary, but setting the temperature $T$ to exactly vanish. The quark chemical potentials $\mu_f$ are set equal (i.e., $\mu_f \equiv \mu = \mu_\text{B}/3$) in anticipation of the beta-equilibrated $N_f=3$ case, but this assumption is straightforward to relax if need be. 

To complete the state-of-the-art order $\alpha_s^3$ in the weak-coupling expansion of the pressure \cite{Gorda:2021znl,Gorda:2021kme,Gorda:2023mkk}, the task still remaining is to evaluate the fully hard contribution $p^h$, which enters the expansion through the four-loop vacuum graphs of the full theory. Applying dimensional regularization in $D\equiv d+1 = 4-2\epsilon$ to regulate both ultraviolet (UV) and IR divergences and renormalization to remove the UV poles, this quantity obtains the form (see \cite{Gorda:2023mkk})
\begin{eqnarray}
    p^\text{h}= \left (\frac{\alpha_s}{\pi} \right )^{\!\!3} \frac{d_A\mu^4}{(4\pi)^2}  \! \biggl[ \frac{p^\text{h}_{-2}}{(2\varepsilon)^2} + \frac{p^\text{h}_{-1}(\Lbar)}{2\varepsilon} + p^\text{h}_0(\Lbar) +O(\varepsilon)\biggl], \,\label{eq:hardpressure}   
\end{eqnarray}
where $\alpha_s$ runs with the $\overline{\text{MS}}$ renormalization scale $\Lbar$ and all remaining poles are of IR origin. The IR-divergences $p^h_{-2}$ and $p^h_{-1}$ and the IR-finite coefficient
\begin{eqnarray}
p^h_0(\Lbar) &=& \sum_{k=1}^{3} c_k(\Lbar)N_f^k, \label{eq:ph0}
\end{eqnarray}
grouped here in powers of $N_f$, are determined from the 52 four-loop diagrams of fig.~\ref{fig:diags}, each containing at least one quark loop. Of these graphs, the 12 diagrams in the leftmost group are all observed to vanish, either because they contain a factorizing scalefree (purely bosonic) subdiagram or because the color trace of the graph evaluates to zero. The other three categories are ordered according to eq.~(\ref{eq:ph0}) and exhibit different IR properties: the largest set of diagrams, containing exactly one fermion loop, is IR safe and contributes only to $p_0^h$, while the $O(N_f^2)$ and $O(N_f^3)$ diagrams each exhibit IR sensitivity \cite{Gorda:2021kme}, giving rise to nonzero $p^h_{-2}$ and $p^h_{-1}$ coefficients in eq.~(\ref{eq:hardpressure}).

\begin{table}[t]
    \centering
    \begin{tabular}{||c|c|c|c|c|c|c|}
       \hline
       \# Ints. & $\;N_f^3\;$ & $\;N_f^2C_A\;$ & $\;N_f^2C_F\;$ & $\;N_fC_A^2\;$ & $\;N_fC_AC_F\;$ & $\;N_fC_F^2\;$ \\
       \hline \hline
       $\xi^0$ & 132 & 2229 & 958 & 5975 & 2841 & 890 \\
       \hline
       $\xi^1$ & 205 & 7428 & 2054 & 34554 & 11507 & 2209 \\
       \hline
       $\xi^2$ & 173 & 9461 & 2452 & 72831 & 17340 & 2949 \\
       \hline
       $\xi^3$ & 125 & 5507 & 1080 & 75344 & 10951 & 1300 \\
       \hline
       $\xi^4$ & - & 2632 & - & 44618 & 3491 & - \\
       \hline
       $\xi^5$ & - & - & - & 20036 & - & - \\
       \hline
       \hline
      $\xi^0$ & 18 & 50 & 48 & 65 & 55 & 45 \\
\hline
    \end{tabular}
    \caption{First six rows: The numbers of distinct integrals in each gauge-invariant sector, proportional to various powers of the gauge parameter $\xi$. Last row: The same numbers after the simplifications described in Appendix \ref{app:reductionalgorithm}, upon which all dependence on $\xi$ vanishes. Note that we have suppressed the common multiplicative factor $d_A\equiv N_c^2-1$ from each invariant and defined $C_A\equiv N_c$, $C_F\equiv(N_c^2-1)/(2N_c)$.}
    \label{tableintegrals}
\end{table}

Upon performing the color traces and Lorentz algebra in a general covariant gauge with the help of \texttt{FORM} \cite{Vermaseren:2000nd}, all four-loop scalar Feynman integrals can be uniquely defined in terms of a fixed list of ten propagators, resulting in a set of 156307 different integrals (see Appendix \ref{app:reductionalgorithm}). At this point, the integrals from the $N_f$ and $N_f^2$ graphs can be further subdivided to three and two respective categories based on the group-theory invariants they contain. The result is displayed in Table \ref{tableintegrals}, where we indicate the numbers of individual terms at this stage of the calculation, sorted by powers of the gauge parameter $\xi$. 

\begin{figure}[t]
    \centering    
    \includegraphics[width=0.48\textwidth]{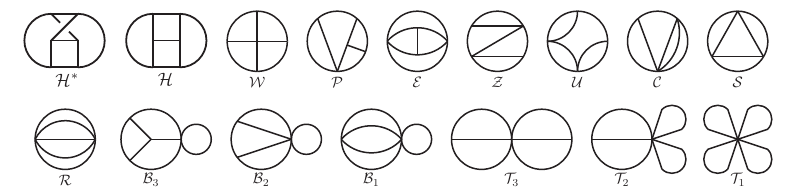}
    \caption{The four-loop vacuum topologies of QCD that produce our 114 master integrals upon assigning bosonic and fermionic signatures for the propagators. See also Appendix \ref{app:reductionalgorithm} and \cite{Navarrete:2024ruu} for further discussion.}
    \label{4loopmastersector}
\end{figure}

Leveraging the integrals' invariance under loop momentum redefinitions, a systematic application of linear momentum shifts allows the elimination of substantial redundancy in the resulting expressions. This canonicalization procedure, further described in \cite{Navarrete:2024ruu} and briefly reviewed in Appendix \ref{app:reductionalgorithm} below, leads to an explicit cancellation of the parameter $\xi$, demonstrating the expected gauge invariance of the pressure up to this order. At the same time, the number of seemingly independent integrals is also seen to collapse, with altogether 114 masters remaining (note that there are redundancies on the last line of Table \ref{tableintegrals}). Each master integral further falls into one of the four-loop topologies displayed in fig.~\ref{4loopmastersector}, with all but the $\mathcal{R}$ topology present in our final result.

Despite the observed reduction in the number of masters, their case-by-case evaluation using existing traditional methods, such as the cutting rules of \cite{Ghisoiu:2016swa}, would be far too laborious a strategy given that our result contains all of most complicated two-particle-irreducible (2PI) topologies. Instead, we propose an alternative strategy, where we start by identifying the IR divergent master integrals through a power-counting prescription, validated by the cancellation of IR poles against their HTL counterparts. This leaves a manageable set of IR-safe integrals to be determined, of which a sizeable subset is directly amenable to the numerical dLTD method.

\section{Infrared sensitive diagrams}

The IR sensitivity observed in four-loop vacuum graphs with either two or three quark loops can be shown to arise from gluon self-energy or vertex-type subdiagrams that do not vanish in the limit of small external momenta. The leading soft behavior of such bosonic $n$-point  graphs is described by the HTL effective theory and is typically considerably simpler than their full expressions, so that the IR divergences of the original vacuum diagrams may be contained in only a few master integrals. This is notably the case for the bugblatter diagrams considered in \cite{Navarrete:2024zgz}, for which only one single master-integral topology --- the $\mathcal{H}$ in fig.~\eqref{4loopmastersector} --- is found to exhibit an IR divergence (see also eq.~\eqref{appeq:IRbugblatter} below).

Inspired by these observations and equipped with the dLTD method capable of numerically handling IR-safe integrals, we will now seek to identify and isolate the IR-sensitive masters, in which we use the canonicalization procedure outlined in Appendix \ref{app:reductionalgorithm} and a power-counting scheme based on the known HTL expressions of the self-energy and vertex subdiagrams. After this, we will explicitly determine the IR divergences of the relevant masters, expecting their sum to cancel against poles identified in the effective-theory contributions evaluated in \cite{Gorda:2021znl,Gorda:2023mkk}. \textcolor{black}{The remaining IR-subtracted integrals will finally be evaluated together with the inherently IR-convergent masters}.

\begin{table}[t]
\centering
\begin{tabular}{||c|c|c|}
    \hline
    Topology & $\widetilde{p}^\text{\,h}_{-2}$ & $\widetilde{p}^\text{\,h}_{-1}$ \\ 
    \hline \hline
    \includegraphics[scale=1.0]{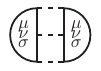} & \raisebox{0.5cm}{$0$} & \raisebox{0.5cm}{$-0.17590(60)C_A$} \\
    \hline
    \includegraphics[scale=1.0]{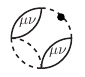} & \raisebox{0.5cm}{$\frac{11}{6}C_A$} & \raisebox{0.5cm}{$\left(-\frac{11}{3}L-3.22027\right)C_A$} \\
    \hline
    \includegraphics[scale=1.0]{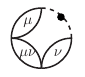} & \raisebox{0.5cm}{$0$} & \raisebox{0.5cm}{$\left(3-\frac{\pi^2}{4}\right)\left(2C_F-C_A\right)$} \\
    \hline
    \includegraphics[scale=1.0]{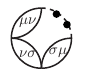} &
    \raisebox{0.5cm}{$0$} & \raisebox{0.5cm}{$\left(\frac{7\pi^2}{144}-\frac{5}{12}+\frac{2}{3}L\right)N_f$} \\
    \hline
    \includegraphics[scale=1.0]{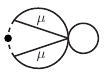} &
    \raisebox{0.5cm}{$0$} & \raisebox{0.5cm}{$\left(8-\frac{2\pi^2}{3}\right)C_F$} \\
    \hline
    \includegraphics[scale=1.0]{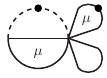} &
    \raisebox{0.5cm}{$0$} & \raisebox{0.5cm}{$-2C_F$} \\
    \hline
    \includegraphics[scale=0.7]{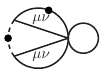} $\!\!\!$\includegraphics[scale=0.7]{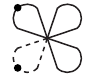} & \multirow{2}{*}{$0$} & \multirow{2}{*}{$\big(-1+\frac{\pi^2}{3}\big)C_F$} \\
    $\!$\includegraphics[scale=0.7]{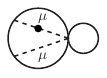} $\!\!\!\!\!$ \includegraphics[scale=0.7]{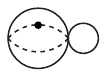} &
     &  \\
    \hline
    \includegraphics[scale=1.0]{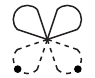} &
    \raisebox{0.5cm}{$0$} & \raisebox{0.5cm}{$-\frac{1}{2}C_A$} \\
    \hline \hline
    \multirow{2}{*}{Total} & \multirow{2}{*}{$\frac{11}{6}C_A$} & \raisebox{0.05cm}{$\left(-\frac{11}{3}L-4.42877(60)\right)C_A+$}\rule{0pt}{3ex} \\
    & & \raisebox{0.15cm}{$\left(\frac{66-5\pi^2}{6}\right)C_F+\left(\frac{7\pi^2}{144}-\frac{5-8L}{12}\right)N_f$}\rule{0pt}{3ex} \\
    \hline
\end{tabular}
\caption{The UV-renormalized contributions of all IR-sensitive four-loop scalar topologies to eq.~\eqref{eq:hardpressure}, following the canonicalization of the pressure of Appendix~\ref{app:reductionalgorithm} and the various manipulations described in Appendix~\ref{app:IR}. Here, the Greek indices reflect the tensor structure of the vertex, bubble, and tadpole insertions, while the solid and dashed lines represent fermionic and bosonic signatures, respectively, and the dotted line a squared propagator. Note that we make use of the short-hand notation $L\equiv\ln\Lbar/(2\mu)$ here and that the results of altogether four topologies have been combined due to the appearance of spurious divergences that cancel in their sum.}
\label{tableIR}
\end{table}

The details concerning the extraction of IR divergences are presented in Appendix \ref{app:IR} below and more extensively in a forthcoming companion paper. 
Given that the $p^\text{h}_{-2}$ and $p^\text{h}_{-1}$ coefficients of eq.~\eqref{eq:hardpressure} are linked to effective-theory divergences, our results automatically contain the thermal screening mass $m_\text{E}^2=(2\alpha_s/\pi)N_f\mu^2 + O(\varepsilon)$, for which an unexpanded $d$-dimensional version is used in the calculations. 
This makes it natural to write the IR divergences in the form \footnote{Note that in the notation of \cite{Gorda:2023mkk}, the regulatory scale parameter in this expression would be the factorization scale $\Lambda_h$ instead of $\Lbar$. 
Given that we have explicitly witnessed the cancellation of $\Lambda_h$ between the full theory and HTL contributions to the pressure, we have identified the two scales here.}
\begin{eqnarray}
    p^\text{h}_{-j}\equiv \frac{\pi^2m_\text{E}^4}{4\mu^4\alpha_s^2}\left(\frac{\mu}{\Lbar}\right)^{-4\varepsilon}\widetilde{p}^\text{\,h}_{-j}(\Lbar),\quad j=1,2,
\end{eqnarray}
where $\widetilde{p}^\text{\,h}_{-2}$ and $\widetilde{p}^\text{\,h}_{-1}$ stand for $\Lbar$-dependent coefficients with values predicted in \cite{Gorda:2021znl,Gorda:2023mkk} based on the form of the HTL results derived therein. The results of our direct evaluation of these coefficients, explained in Appendix~\ref{app:IR}, are shown in table~\ref{tableIR}. Upon comparison, we observe perfect agreement of the sum of the IR divergences with eq.~(42) of \cite{Gorda:2023mkk}, confirming their full cancellation from the $O(\alpha_s^3)$ pressure of cold and dense QM.

As explained in \cite{Gorda:2023mkk}, the $\Lbar$ dependence of the $c_k(\Lbar)$ coefficients, defined in eq.~\eqref{eq:ph0} above, can be analytically determined from renormalization-scale independence of the full pressure, leaving pure numbers to be computed. For the contribution proportional to the maximal power of the number of quark flavors, $c_3$, even the full result is known and agrees with the Abelian case of \cite{Gorda:2022zyc} up to an overall color factor. The result reads
\begin{equation}
c_3(\Lbar) =  0.101515 + 0.805957 \ln \frac{\Lbar}{2\mu} + \frac{8}{9}\left(\ln\frac{\Lbar}{2{\mu}}\right)^{2},
\end{equation}
where the numerical coefficients can be evaluated to high precision \cite{Gorda:2022zyc}. This leaves only a set of IR-safe masters, contributing to $c_1$ and $c_2$, to be determined.

\section{Infrared safe sector}\label{sec:IRsafe}

Moving on to the IR-safe sector, which fully determines the $c_1(\Lbar)$ coefficient in eq.~\eqref{eq:ph0}, we will next outline the application of the dLTD method to two non-factorizing four-loop diagrams. Importantly, these graphs can be evaluated using traditional semi-analytic methods as well \cite{Kurkela:2009gj,Ghisoiu:2016swa}, enabling the first-ever direct tests of the accuracy of the numerical dLTD method at four-loop order.

The dLTD procedure begins with the application of the Bogoliubov $R$-formula to the original diagram to ensure the local finiteness of the integrand in the eventual numerical integrations \cite{Bogoliubov:1957gp,Hepp:1966eg,Zimmermann:1969jj}. In practice, this amounts to algorithmically subtracting a set of counterterms $\mathrm{CT}_i[\Gamma]$ from the original graph $\Gamma$ in  $D=4-2\varepsilon$ dimensions, which has the effect of removing all UV divergences including nested and overlapping ones,
\begin{equation}
\begin{split}
    \Gamma &= \underbrace{\left(\Gamma - \sum_i\mathrm{CT}_i[\Gamma]\right)}_{D=4} + \underbrace{\sum_i\mathrm{CT}_i[\Gamma]}_{D\neq 4} \\
    &\equiv R[\Gamma] + \sum_i\mathrm{CT}_i[\Gamma]\, .\label{eq:UVschematic}
\end{split}
\end{equation}
Here, $R[\Gamma]$ represents a UV-finite integral amenable to numerical treatment in $d=3$ after the analytic evaluation of the temporal momentum integrals, while the UV-divergent counterterms $\mathrm{CT}_i[\Gamma]$ are evaluated analytically in $D=4-2\varepsilon$ using the library \texttt{vakint} \cite{vakintgit}. For the first time ever, including vacuum applications, we have fully automated this procedure in momentum space up to arbitrarily high loop orders.

To benchmark the dLTD method against traditional ones, we select the first two $O(\nf)$ diagrams from fig.~\ref{fig:diags}, which take relatively simple forms after the canonicalization procedure of Appendix \ref{app:reductionalgorithm}. A straightforward calculation produces
\begin{align}
    -\frac{1}{12}\raisebox{-0.4\height}{\includegraphics[height=1.2cm]{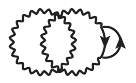}}&=-\frac{3}{4}d_A N_f C_A^2 g^6 d(d-1) I_\mathcal{C},\label{eq:nfdiag1} \\
    -\frac{1}{8}\raisebox{-0.4\height}{\includegraphics[height=1.2cm]{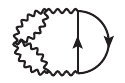}}&=\frac{27}{16} d_AN_f C_A^2 g^6 d(d-1) I_\mathcal{S},\label{eq:nfdiag2}
\end{align}
where the IR-safe but UV-divergent master integrals corresponding to  the topologies $\mathcal{C}$ and $\mathcal{S}$ of fig.~\ref{4loopmastersector} read
\begin{eqnarray}
    I_\mathcal{C} &\equiv& \int_{PQR\{S\}}\frac{1}{P^2 Q^2 R^2 S^2 (P-S)^2 (P-Q-R)^2}, \label{eq:I1int} \\
    I_\mathcal{S} &\equiv& \int_{\{P\}QRS}\frac{1}{P^2 Q^2 R^2 (P-S)^2 (Q-S)^2 (R-S)^2} , \;\;\;\;\label{eq:I2int}
\end{eqnarray}
with integration measures defined in eqs.~(\ref{C1})-(\ref{C3}). 

Before engaging in a numerical dLTD calculation, a few words of warning are in order. Should one attempt to apply the procedure directly to the graphs in eqs.~(\ref{eq:nfdiag1}) and (\ref{eq:nfdiag2}), the squared gluon propagators separating the one- and two-loop self energy structures would namely lead to \emph{spurious} IR divergences, only canceling upon the evaluation of the medium-independent two-loop self-energies. To prevent this from causing problems in numerical calculations, one possibility described in \cite{Capatti:2022tit} is to modify the UV-subtraction operators introduced above. Here we, however, take advantage of the IR finiteness of the master integrals in eqs.~(\ref{eq:I1int}) and (\ref{eq:I2int}), a feature common to the entire $N_f$ sector.

\begin{table}[t]
\centering
\begin{tabular}{||c|c|c|c|c|c|c|}
    \hline
    Diagram & $\varepsilon^{-2}$ & $\varepsilon^{-1}$ & $\varepsilon^0_\text{traditional}$ & $\varepsilon^0_\text{dLTD}$ & $N\,[10^6]$ & $[\mathrm{\mu s}]$ \\ 
    \hline \hline
    \raisebox{1.3\height}{$-\frac{1}{12}$}\includegraphics[height=0.9cm]{4loopdiags/diag23.pdf} & \raisebox{1.3\height}{$0$} & \raisebox{1.3\height}{$\frac{3}{4}$} & \raisebox{1.3\height}{$12.375$} & \raisebox{1.2\height}{$12.36(4)$} & \raisebox{1.3\height}{$110$} & \raisebox{1.3\height}{$7.1$} \\
    \noalign{\hrule height.9pt} \noalign{\hrule height.9pt}
    \raisebox{1.3\height}{$-\frac{1}{8}$}\includegraphics[height=0.9cm]{4loopdiags/diag111.pdf} & \raisebox{1.3\height}{$\frac{27}{4}$} & \raisebox{1.3\height}{$\frac{189}{2}$} & \raisebox{1.3\height}{$716.38$} & \raisebox{1.2\height}{$716.32(7)$} & \raisebox{1.3\height}{$120$} & \raisebox{1.3\height}{$6.5$} \\
    \hline
\end{tabular}
\caption{First terms in the $\varepsilon$ expansions of the two diagrams in eqs.~(\ref{eq:nfdiag1}) and (\ref{eq:nfdiag2}), omitting an overall prefactor of $d_AN_fC_A^2g^6d(d-1)\mu^4/(4\pi)^8$ and setting $\Lbar=2\mu$ (see Appendix~\ref{app:analyticalComputation} for $\Lbar$-dependent results). Here, $N$ indicates the number of Monte Carlo samples needed and $[\mathrm{\mu s}]$ the time spent on one sample on a single CPU core.}
\label{tableNumericalValues}
\end{table}

The results of our practical calculations are summarized in table~\ref{tableNumericalValues}. 
Employing the \texttt{vegas} integration routine \cite{vegasgit} with approximately $10^8$ Monte Carlo samples, the dLTD computation leads to numerical results that agree with the analytical values obtained with cutting rules, discussed in  Appendix~\ref{app:analyticalComputation}, to sub-percent precision. The table also lists the Monte Carlo statistics and per-sample-per-core timings in microseconds, translating to evaluation times of just a few minutes on a standard quad-core laptop. In comparison with the two- and three-loop benchmarks discussed in \cite{Navarrete:2024zgz}, these results showcase the efficient scalability of the method to higher orders.

\section{Discussion and outlook}

Determining the pressure of cold and dense quark matter to the full $\alpha_s^3$ order is a longstanding problem in perturbative thermal field theory with important phenomenological applications in the physics of neutron stars. Completing the task requires, however, overcoming formidable theoretical and technical challenges related to the evaluation of four-loop vacuum diagrams in the limit of vanishing temperature but nonzero quark chemical potentials. The more conceptual challenges concern the cancellation of infrared divergences from the weak-coupling expansion of the pressure, while the technical challenge lies in the extremely demanding evaluation of the remaining IR-finite integrals.

In this paper, we have taken decisive steps to overcoming the issues listed above. By systematically inspecting the sum of all four-loop vacuum graphs of dense QCD, we not only witnessed the full cancellation of the covariant gauge parameter, but also demonstrated the IR-finite nature of the pressure. The latter was achieved by identifying the IR sensitive four-loop graphs, extracting the IR divergences therein, and witnessing their exact cancellation against contributions from the Hard Thermal Loop effective theory, previously determined in \cite{Gorda:2021znl,Gorda:2023mkk}.

Being left with a sizable set of IR-finite integrals, we demonstrated the applicability of a novel computational tool, the numerical dense Loop Tree Duality method, for the problem. We chose two non-factorizable four-loop graphs, amenable to evaluation with traditional methods but presenting a challenging UV structure for dLTD, and found perfect agreement between the results. This is an important result, given that the traditional methods are not merely cumbersome but downright inapplicable to many of the more complicated topologies, while the dLTD method allows a high degree of automation and is largely insensitive to the structure of the Feynman graph.

With the sole $O(N_f^3)$ graph of fig.~\ref{fig:diags} and two of the $O(N_f)$ diagrams now fully determined and the IR issues resolved, the way forward is in principle clear. Although still a formidable task, the systematic application of the dLTD method will eventually lead to the successful numerical evaluation of the remaining IR safe integrals, thus completing the determination of the $O(\alpha_s^3)$ pressure of cold and dense QM. Alongside this, the dLTD formalism is already being generalized to various directions, including nonzero temperatures, Feynman graphs with external legs, and even the real-time formalism of thermal field theory. Indeed, this method carries great potential to overcome challenges previously considered untractable, and its applicability extends far beyond the realm of cold and dense systems.

\section*{Acknowledgements}
We thank Zeno Capatti, Lo\"ic Fernandez, Mathijs Fraaije, Tyler Gorda, Valentin Hirschi, Lucien Huber, Aleksi Kurkela, York Schröder, Saga Säppi, and Juuso Österman for useful discussions. Our work has been supported by the Research
Council of Finland grants 347499, 353772, 354533, and 354572, as well as the Magnus Ehrnrooth foundation. Computational resources have been provided by CSC - IT Center for Science, Finland, and all figures have been prepared with Axodraw \cite{Collins:2016aya}.

\bibliography{references.bib}

\section*{End matter}

\appendix

\section{Reduction algorithm and integral basis}\label{app:reductionalgorithm}

In this first appendix, we describe the symbolic manipulations that reduce the $O(\alpha_s^3)$ pressure to its canonical form through a finite-$\mu$ generalization of an in-house code developed for \cite{Navarrete:2024ruu} and further used in \cite{Navarrete:2024zgz}. Readers interested in the technical details of the code are referred to Section III of \cite{Navarrete:2024ruu}.

Working in the limit of massless quarks, we can write all propagators appearing in the four-loop vacuum diagrams in terms of squared momenta $P_i^2\equiv P_i\cdot P_i$, with $\{P_i\}_{i=1}^{10}$ constructed from linear combinations of the four independent loop momenta $\{K_j\}_{j=1}^4$,
\begin{align}
    P_i = \sum_{j=1}^4 \lambda_{ij}K_j. \label{app:momfamily}
\end{align}
Upon a specific choice of the $\lambda_{ij}$, and thereby $\{P_i\}$, each four-loop scalar vacuum integral is expressed as a collection of ten numbers $a_i\in \mathds{Z}$, with a positive value representing the power of the corresponding propagator $(P^2_i)^{-1}$ and a negative value signaling the presence of a numerator in the form of an inverse propagator. In a finite-density system with one independent chemical potential $\mu$, an additional list of four numbers $s_j$, chosen from the ternary $\{-1,0,+1\}$, is finally needed to specify the signatures of the loop momenta $K_j=(k^0_j+is_j \mu,\mathbf{k}_j)$. Here, $s_j=0$ corresponds to a bosonic propagator and $s_j=\pm1$ defining the direction of fermion flow. 

The above choices completely fix the signature of each scalarized graph, allowing us to express any four-loop finite-density vacuum integral $I$ as a list of 14 integers,
\begin{align}
    I(s_j;a_i)\equiv \int_{\{K_j\}_{s_j}}\frac{1}{(P_1^2)^{a_1}\dots(P^2_{10})^{a_{10}}}. \label{appeq:listnotation}
\end{align}
The pipeline needed to translate the vacuum diagrams of fig.~\ref{fig:diags} to scalarized structures of the form eq.~\eqref{appeq:listnotation} is identical to that used in \cite{Navarrete:2024ruu}. The generation of the diagrams follows from gluing together all connected three-loop gluon self energies with \texttt{Qgraf} \cite{Nogueira:1991ex}, while the insertion of Feynman rules in a general covariant gauge and the subsequent algebraic manipulations are carried out with \texttt{FORM} \cite{Vermaseren:2000nd}. At this stage, the obtained scalar integrals are mapped to lists defined by eq.~\eqref{appeq:listnotation} upon a choice of the momentum family $\{P_i\}$, producing a large set of distinct integrals $\{\tilde{I}_k\}_{k=1}^{\tilde{N}}$, $\tilde{N}=156307$, appearing in the sectors of Table~\ref{tableintegrals}.

With the integrals now written in a convenient form and the pressure accordingly sorted, our reduction algorithm builds upon a systematic use of shifts in the loop momentum variables, introduced in \cite{Navarrete:2024ruu}. Keeping track of changes in propagator signatures induced by these shifts, we systematically map the set $\{\tilde{I}_k\}_{k=1}^{\tilde{N}}$ to a representative class of integrals uniquely defined by a lexicographic ordering among the indices $(s_j;a_i)$ and graphically represented in fig.~\ref{4loopmastersector}. 
The final result is a gauge-independent basis $\{I_k\}_{k=1}^N$ of ``master integrals" expressed in the canonical form of \cite{Navarrete:2024ruu}, with the total number of integrals independent with respect to reparametrizations of loop momenta being $N=114$.

\section{Infrared divergences}\label{app:IR}

As noted above, the scalarization of the diagrams and the further reduction of the resulting expressions allow us to map the IR divergences of the four-loop vacuum graphs onto a small fraction of master integrals within the integral basis $\{I_k\}_{k=1}^{114}$. These integrals typically feature somewhat simpler structures than the full diagrams, which in particular reduces the problem of extracting IR divergences to considering lower-dimensional integrals over subdiagrams expanded in soft external momenta.

To identify  the subset of master integrals that contain IR poles, we first expand all inverse propagators in scalar products of loop momenta, which allows us to express the resulting integrands in terms of contractions of one- and two-loop tensor vertex, bubble, and tadpole substructures. Together with tensor decomposition and a power-counting analysis, knowledge of their behavior in a soft-external-momentum expansion allows us to single out three distinct categories of IR-divergent integrals.

\noindent \textbf{Category 1:} The integrals in the first category include a single momentum integration over a bosonic propagator with an exponent higher than unity, i.e.~take the form
\begin{equation}
    \int_Q \frac{1}{(Q^2)^s}f(Q). \label{appeq:IRrings}
\end{equation}
Here, $f(Q)$ is an IR-safe (though possibly UV-divergent) function of the external bosonic momentum $Q$, built from contractions of several one-loop bubbles or a single two-loop bubble insertion. The former structures are
\begin{align}
    \hat{\Pi}(Q)&\equiv \int_{\{K\}}\frac{1}{K^2({K-Q})^2}, \label{appeq:scalarpi} \\
    \hat{\Pi}^\mu(Q)&\equiv \int_{\{K\}}\frac{K^\mu}{K^2({K-Q})^2}, \label{appeq:vectorpi} \\
    \hat{\Pi}^{\mu\nu}(Q)&\equiv \int_{\{K\}} \frac{2K^\mu K^\nu-K^2\delta^{\mu\nu}}{K^2(K-Q)^2}, \label{appeq:tensorpi} \\
    \widetilde{\Pi}^{\mu\nu}(Q)&\equiv \int_{\{K\}} \frac{2K^\mu K^\nu-K^2\delta^{\mu\nu}}{(K^2)^2(K-Q)^2}, \label{appeq:tensorpitilde}
\end{align}
while the two-loop structures correspond to
\begin{align}
     \Pi_1(Q) &= \int_{\{KR\}} \frac{1}{K^2R^2(K-R+Q)^2}, \label{appeq:Pi1_2loop} \\
     \Pi_2(Q) &= \int_{\{KR\}} \frac{K\cdot R}{K^2 R^2 (K-Q)^2 (K-R-Q)^2}. \label{appeq:Pi2_2loop}
\end{align}
Inspecting our set of \textcolor{black}{masters, we find several integrals with the topologies $\mathcal{U}$, $\mathcal{B}_2$, and $\mathcal{T}_2$ of fig.~\ref{4loopmastersector}} that take the form given in eq.~\eqref{appeq:IRrings}. 

Among the integrals with $s=2$, the IR divergences \textcolor{black}{can be isolated by neglecting terms sub-leading in powers of $Q$  in the function $f(Q)$}. For eqs.~\eqref{appeq:vectorpi} and \eqref{appeq:tensorpi}, only the matter (vacuum-subtracted) parts contribute, giving rise to the $Q\ll\mu$ limits \footnote{The full expressions and corresponding derivations will be given in a forthcoming companion paper.}
\begin{align}
    \hat{\Pi}^\mu(Q)&\overset{\text{LO}}{\sim}i\Pi^\text{L}_\text{HTL}(x) \frac{Q^2U^\mu-Q^0Q^\mu}{|\mathbf{q}|^2},  \\
    \hat{\Pi}^{\mu\nu}(Q)&\overset{\text{LO}}{\sim}\Pi^{\mu\nu}_\text{HTL}(x), \label{appeq:tensorpiHTL}
\end{align}
where $x\equiv Q^0/|\mathbf{q}|$, L denotes the longitudinal part of the HTL self energy $\Pi^{\mu\nu}_\text{HTL}$ (see \cite{Gorda:2021kme}), and $U^\mu=(1,\mathbf{0})$. The HTL functions appearing here are proportional to the screening mass parameter $m_\text{E}^2=(2\alpha_s/\pi)N_f\mu^2 + O(\varepsilon)$, setting the natural scale of the HTL effective theory (see \cite{Gorda:2021kme} for the $d$-dimensional expression). 

In contrast to the above cases, the dimensionless nature of the scalar bubble in eq.~\eqref{appeq:scalarpi} implies that both its matter and vacuum parts source IR-divergent integrals, with the latter being linked to gauge coupling renormalization in effective-theory and lower-loop contributions. These integrals turn the mixed $1/(\varepsilon_\text{IR}\varepsilon_\text{UV})$ poles of the hard sector into double IR poles, present in only one single topology featured on the second row of table~\ref{tableIR}. 

Moving on to the tensor bubble of eq.~\eqref{appeq:tensorpitilde}, appearing on the seventh row of table~\ref{tableIR}, we note that the presence of a squared fermion propagator slightly complicates taking the HTL limit. This is due to the non-interchangeability of the order of the temporal and spatial integrations \cite{Gorda:2022yex}, introducing additional terms in the application of the residue theorem. We note in passing the appearance of an additional squared propagator on the sixth row of table~\ref{tableIR}, featuring the tadpole
\begin{align}
    \int_{\{K\}}\frac{K^\mu}{(K^2)^2}=U^\mu\int_{\{K\}}\frac{K^0}{(K^2)^2}.
\end{align}
This integral results in a purely imaginary contribution that is straightforward to determine by direct integration.

\textcolor{black}{For graphs featuring two-loop bubbles, cf.~eqs.~\eqref{appeq:Pi1_2loop}-\eqref{appeq:Pi2_2loop}, the IR divergences can be determined without relying on $Q\ll\mu$ expansions. For a $\Pi_1$ insertion, the $Q$ integral can be done analytically, reducing the problem to a simple two-loop computation. The IR sensitive pieces of $\Pi_2$ are on the other hand seen to map to $\Pi_1$ and lower-loop-order terms upon a suitable change of variables.}

Returning to eq.~(\ref{appeq:IRrings}), the $s=3$ case is finally specific to the sole $O(N_f^3)$ diagram of fig.~\ref{fig:diags}, requiring the NLO term in the expansion of $\hat{\Pi}^{\mu\nu}(S)$ in soft external momenta, dubbed power corrections in \cite{Gorda:2021znl,Gorda:2021kme}. In this case, we find it more convenient to write one of the $\hat{\Pi}^{\mu\nu}(S)$ insertions in terms of the full self-energy $\Pi^{\mu\nu}(S)$ (see the fourth graph in table~\ref{tableIR}). For this quantity, we get
\begin{align}
    \Pi^{\mu\nu}(Q)=\Pi^{\mu\nu}_\text{HTL}(x)+\frac{\alpha_sQ^2}{4\pi}\Pi^{\mu\nu}_\text{Pow}(x)+O(Q^4),
\end{align}
where $\Pi^{\mu\nu}_\text{Pow}$ is known analytically \cite{Gorda:2023zwy}. Obtaining the IR pole then amounts to performing a straightforward angular integration upon replacing the other two tensor-bubble insertions as in eq.~\eqref{appeq:tensorpiHTL}, leading to the result reported on the fourth row of table~\ref{tableIR}.

\noindent \textbf{Category 2:} The second category of IR-sensitive integrals derives from the 'bugblatter' diagrams and amounts to two masters of the $\mathcal{H}$ topology, adding up to 
\begin{equation}
\int_{PQR} \frac{\delta(P+Q+R)}{P^2Q^2R^2} [\text{Re}\,\hat{\Gamma}^{\mu\nu\sigma}(P,Q,R)]^2. \label{appeq:IRbugblatter}
\end{equation}
This expression involves the real part of the tensor vertex
\begin{align}
    \hat{\Gamma}^{\mu\nu\sigma}(P,Q,R)&\equiv \int_{\{K\}}\frac{K^\mu K^\nu K^\sigma}{K^2(K-R)^2(K+Q)^2}, \label{appeq:IRgamma}
\end{align}
the leading soft behavior of which corresponds to the well-known HTL vertex function \cite{Gorda:2021kme}
\begin{align}
    \hat{\Gamma}^{\mu\nu\sigma}(P,Q,R)\overset{\text{LO}}{\sim}\Gamma^{\mu\nu\sigma}_\text{HTL}(P,Q,R). \label{appeq:HTLgamma}
\end{align}
This function is  real-valued and scales with the magnitudes of its external momenta as $\Gamma^{\mu\nu\sigma}_\text{HTL}(P,Q,R)\sim m_\text{E}^2/|P|$, giving rise to a logarithmic IR divergence in eq.~\eqref{appeq:IRbugblatter}. The resulting pole can be obtained numerically by means of a two-fold integration over Euclidean angles, cf.~the first row of table~\ref{tableIR}. Interestingly, the uniqueness of this particular topology enables a direct matching of eq.~\eqref{appeq:IRbugblatter} with a similar expression in the HTL effective theory \cite{Gorda:2021kme}, allowing us to witness the cancellation of the IR divergence at the integrand level.

\noindent \textbf{Category 3:} Finally, the third and last category of IR divergences originates from scalefree integrals that vanish in dimensional regularization and are therefore typically ignored. Given that this involves a cancellation between UV and IR poles in $D=4-2\varepsilon$ dimensions and that our goal is to observe the vanishing of full-theory IR divergences against effective-theory contributions, we need to be more careful in our analysis and keep track of logarithmically IR divergent scalefree integrals.

\textcolor{black}{The IR-sensitive scalefree one- and two-loop structures we are after arise from master integrals with the topologies $\mathcal{T}_1$ and $\mathcal{T}_2$. With our aim being to witness an exact cancellation of the hard IR divergences against HTL contributions, the corresponding integrals are evaluated keeping their UV and IR divergences distinct, following simplifications using IBP techniques.}

\textcolor{black}{In the $C_F$ sector, scalefree integrals turn out to play an important role in removing spurious double poles present in graphs that mix UV and IR divergences, leading to the IR pole contribution on the seventh row of table~\ref{tableIR}. Meanwhile, scalefree contributions in the $C_A$ group can be traced back to a single four-loop graph, the first diagram in the third category of fig.~\ref{fig:diags} (corresponding to the HTL-resummed $I_{4g}$ in \cite{Gorda:2021kme}). Once this connection is established, the scalefree hard integrals can be matched against the corresponding effective-theory terms that turn out to feature trivial angle dependence. The result of this exercise is visible on the second-to-last row of table~\ref{tableIR}.}

Finally, we stress that keeping UV and IR divergences distinct in our calculation serves solely as a means to show the exact cancellation of IR divergences. In practice, the full pressure computation can be carried out consistently with the help of a single regulator $\varepsilon$.

\section{Details of the cutting-rule calculation}\label{app:analyticalComputation}

In the last of our three appendices, we provide details of our analytic determination of the IR-safe four-loop integrals $I_{\mathcal{C}}$ and $I_{\mathcal{S}}$ of eqs.~\eqref{eq:I1int} and \eqref{eq:I2int} using the cutting-rule method introduced in \cite{Ghisoiu:2016swa}.
In the calculation, we employ the bosonic integration measure
\begin{equation}
\int_{S} \equiv \left (\frac{e^{\gamE}\Lbar^2}{4\pi} \right )^{\frac{4-D}{2}} \int \frac{\ud^D S}{(2\pi)^D} = \int_{-\infty}^{+\infty} \frac{\ud s_0}{2\pi} \int_{\st},  \label{C1}
\end{equation}
where $\Lbar$ is the $\overline{\text{MS}}$ renormalization scale, $\gamE$ the Euler–Mascheroni constant, and we have introduced the common short-hand for spatial integrals,
\begin{equation}
\begin{split}
\int_{\st} & \equiv \left (\frac{e^{\gamE}\Lbar^2}{4\pi} \right )^{\frac{3-d}{2}} \int \frac{\ud^d \st}{(2\pi)^d}.
\end{split}
\end{equation}
Similarly, fermionic integrals are denoted as
\begin{align}
\int_{\{S\}}f(s_0,\st)=\int_Sf(s_0+i\mu,\st), \label{C3}
\end{align}
and phase-space integrals, associated with the cutting-rule method, are defined according to
\begin{equation}
\cint_{\!\!\!\!\!\:\st}~\equiv \int_\st \frac{\theta(\mu-s)}{2s}, \; s \equiv \vert \st \vert.
\end{equation}

We begin with the integral  $I_{\mathcal{C}}$. Upon the change of variables $P\rightarrow S-P$, eq.~\eqref{eq:I1int} becomes
\begin{equation}
  I_{\mathcal{C}} = \int_{\{PS\}QR}\frac{1}{(S-P)^2Q^2R^2S^2P^2(S-P-Q-R)^2}, \label{C1000Integral}
\end{equation}
to which we apply the cutting rules. This leads to the schematic expression
$I_{\mathcal{C}}  = \sum_{i=0}^2 I_{\mathcal{C}}^{i-\text{cut}},$
where the upper limit of the sum is set by the number of linearly independent fermion propagators in the graph. Discarding the zero-cut term as a (vanishing) contribution independent of $\mu$, we are left with only two terms, of which the one-cut contribution vanishes, too, as proportional to the on-shell ($P^2=0$) limit of a scalefree two-point function (see \cite{Ghisoiu:2016swa} for details). This leaves as the only nonvanishing cut 
\begin{equation}
   I_{\mathcal{C}}^\text{2-cut}= \;\:\cint_{\!\!\!\!\!\:\pt\st}\Bigg[\frac{1}{(P-S)^2}\int_{QR} \frac{1}{Q^2R^2(Q+R+P-S)^2}\Bigg],\!\!\!
\end{equation}
where the cut momenta $P$ and $S$ are placed on shell by setting $P=(ip,\pt)$ and $S=(is,\st)$.\;\;

The integrations over $Q$ and $R$ can now be performed analytically using the standard vacuum result 
\begin{eqnarray}
    \int_Q \frac{1}{Q^{2\alpha} (Q - L)^2}&=&\frac{A_{\alpha}(\varepsilon, \Lbar^2)}{(L^2)^{\varepsilon + \alpha - 1}}, \\
A_{\alpha}(\varepsilon,\Lbar^2) &=& \frac{1}{(4\pi)^{2-\varepsilon}} \left (\frac{e^{\gamE}\Lbar^2}{4\pi}\right )^{\varepsilon}\\
&\times&\frac{\Gamma(\alpha + \varepsilon -1)}{\Gamma(\alpha)} \frac{\Gamma(1-\varepsilon)\Gamma(2-\alpha-\varepsilon)}{\Gamma(3-\alpha-2\varepsilon)}.     \nonumber
\end{eqnarray}
This yields the result
\begin{equation}
   I_{\mathcal{C}}= A_1(\varepsilon,\Lbar^2)A_\varepsilon(\varepsilon,\Lbar^2)\cint_{\!\!\!\!\!\:\pt\st}(-2P\cdot S)^{-2\varepsilon}, \label{Ic1}
\end{equation}
where $-2P\cdot S = 2ps(1-\pt\cdot \st/(ps)) \geq 0$. 

Luckily, the radial and angular integrals remaining in eq.~(\ref{Ic1}) above factorize, allowing us to perform them separately. This results in the expression
\begin{align}
   I_{\mathcal{C}}=-\frac{\mu^4}{(4\pi)^8}\Bigg\{\frac{1}{\varepsilon}+8\ln{\frac{\Lbar}{2\mu}+\frac{33}{2}}\Bigg\}+O(\varepsilon),\label{Iaresult}
\end{align}
while a highly analogous computation for integral $I_{\mathcal{S}}$ in eq.~\eqref{eq:I2int} produces
\begin{equation}   
\begin{split}    
I_{\mathcal{S}}&=\frac{4{\mu}^{4}}{(4{\pi})^{8}}\Bigg\{\frac{1}{{\varepsilon}^{2}}+\frac{1}{\varepsilon}\left(14+8\ln\frac{\Lbar}{2{\mu}}\right) +116-{\pi}^{2} \\
   &+112\ln\frac{\Lbar}{2{\mu}}+32\left(\ln\frac{\Lbar}{2{\mu}}\right)^{2}\Bigg\}+O(\varepsilon).
   \label{Ibresult}
\end{split}
\end{equation}
It is worth noting that we were able to evaluate both $I_\mathcal{C}$ and $I_\mathcal{S}$ in a closed form in terms of Euler $\Gamma$ functions, with the $\varepsilon$-expansions shown above performed on these expressions. Substituting $\Lbar=2\mu$ in eqs.~\eqref{Iaresult} and \eqref{Ibresult} finally produces the numerical values shown in the $\epsilon^0_\text{traditional}$ column of table \ref{tableNumericalValues}, in perfect agreement with the dLTD results.

\end{document}